\documentclass[
    ,final            
  ]
  {aipproc}
\layoutstyle{6x9}

\begin{document}

\title{Accounting for the Unresolved X-ray Background with Sterile Neutrino Dark Matter}


\author{D.T.~Cumberbatch}{
  address={Department of Astrophysics, University of Oxford, Keble Road, Oxford, OX1 3RH}
}

\author{Joseph Silk}{
  address={Department of Astrophysics, University of Oxford, Keble Road, Oxford, OX1 3RH}
}

\begin{abstract}
We consider a scenario where keV sterile neutrinos constitute all of the currently inferred dark matter abundance, whose radiative decays could potentially account for the flux contributions to the X-ray background (XRB) by unresolved sources. Here we apply \emph{integrated flux} methods to results from the observations of the North/South \emph{Chandra deep fields} (CDF-N/S) in order to deduce constraints on the sterile neutrino mass-mixing parameters.
\end{abstract}

\maketitle


\section{introduction}
The proportion of the cosmological energy density manifested as dark matter is now well determined by observations to within 10\% \cite{Spergel:2006hy}, yet its nature remains a subject of extensive debate.
Recent astrophysical observations, such as indications of central cores in low-mass
galaxies (see e.g.~\cite{Dalcanton:2000hn}), and the deficiency of satellites observed in Milky Way-sized galaxies (see e.g.~\cite{Klypin:1999uc}), indicate possible shortcomings of the [$\Lambda$]CDM paradigm and have boosted interest in a warm dark matter (WDM) scenario which may
alleviate these ``small scale problems'' (see e.g.~\cite{Bode:2000gq}). 
One of the most popular WDM candidates are right-handed 
( ``sterile'' ) neutrinos $\nu_s$, since they naturally arise in many extensions of the Standard Model (see e.g.~\cite{Asaka:2005an}).
Other motivations for sterile neutrinos to possess mass-mixing parameters necessary for them to be produced via oscillations (with the familiar ``active'' neutrino species) include the enhanced production of neutral hydrogen before reionization~\cite{Biermann:2006bu} and a possible explanation for the baryon asymmetry of the Universe~\cite{Asaka:2005an}.

Sterile neutrinos can be produced through non-resonant oscillations, the simplest being the popular Dodelson-Widrow (DW) mechanism~\cite{Dodelson:1993je}, or resonant oscillations for cosmologies which include a finite lepton number asymmetry. In both scenarios, sterile neutrinos are predominantly produced during the QCD epoch. Unfortunately, uncertainties relating to hadronic interactions during this time result in uncertainties in the relationship between the mass-mixing parameters and the relic abundance of sterile neutrino dark matter \cite{Asaka:2005an, Asaka:2006rw}.

{\em Direct} constraints on the mass-mixing parameters can be obtained by exploiting 
X-ray observations since keV-mass sterile neutrinos posses a radiative decay
channel, yielding photons potentially detectable in astrophysical X-ray sources, 
including the diffuse cosmic X-ray background~\cite{Boyarsky:2005us}, 
the Milky Way~\cite{Boyarsky:2006ag, Abazajian:2006jc} as well as being inferred from the
X-ray spectra of nearby galaxies~\cite{Watson:2006qb} 
or clusters (see e.g.~\cite{Riemer-Sorensen:2006pi}). 
{\em Indirect} constraints can be achieved from
Ly-{$\alpha$} forest measurements~\cite{Narayanan:2000tp}, 
since they are sensitive tracers of the primordial density fluctuations on the
smallest scales where WDM typically suppresses clustering.

Assuming that sterile neutrinos constitute all of the dark matter 
and are produced solely through the DW mechanism, 
recent X-ray analyses (see e.g.~\cite{Boyarsky:2005us,Abazajian:2006jc,Boyarsky:2006ag, Watson:2006qb,Riemer-Sorensen:2006pi})
deduce upper limits on their mass, $m_s\simeq[3-8]$~keV. 
Conversely, the latest Ly-$\alpha$ analyses~\cite{Seljak:2006qw} of
the high redshift power spectra from the SDSS
~\cite{McDonald:2004eu} furnish constraints $m_s<(10-13)$~keV.
Hence, these results seem to exclude scenarios where all dark matter 
consists of sterile neutrinos produced through the DW mechanism. (However, such evidence does not preclude the possibility of a sub-dominant sterile neutrino dark matter component~\cite{DTC_sub_dominant}.)
However, hidden systematic effects within SDSS measurements of the Ly-$\alpha$ flux power spectrum could relax the severity of the corresponding constraints inferred from the 1-D matter power spectrum from the Ly-$\alpha$ forest \cite{Abazajian:2006jc}.

In these proceedings we apply an \emph{integrated flux} method 
to the CDF-N/S spectra of the XRB in order to constrain the radiative decay rate of DW sterile neutrino dark matter.

\section{Flux from sterile neutrino decay} 
Sterile neutrinos possess a radiative decay channel
into an active neutrino and a photon with energy $E_{\gamma} = m_s/2$. 
For Majorana neutrinos, the radiative decay rate~\cite{Pal:1981rm, Barger:1995ty} can be expressed as
\begin{equation}
\Gamma_\gamma\simeq 1.38 \times 10^{-22} \sin^2 2\theta_s
\left(\frac{m_s}{\mathrm{1\,keV}}\right)^5 \mathrm{s}^{-1} \,,
\label{eq:Gamma}
\end{equation}
where $\theta_s$ is the mixing angle in vacuum between the
sterile and the active neutrino species. (For Dirac neutrinos the above rate is halved~\cite{Barger:1995ty}.)

The predicted signal from such decays
originates from both extra-galactic (EG) neutrinos, as well as those residing in
the Milky Way (MW) halo. The EG contribution can be evaluated
assuming a uniform distribution of neutrinos in the visible
Universe up to  very small redshifts. The
differential energy flux (energy flux per unit energy, integrated over the field 
of view (FOV), $\Omega_{\mathrm{FOV}}$, of the detector ) can be 
expressed as~\cite{Masso:1999wj}
\begin{equation}
\varphi_E^{\mathrm {EG}}\simeq \frac{\Gamma_\gamma}{4\pi m_s}
\frac{\Omega_{\mathrm {dm}}~\rho_{c}~\Omega_{\mathrm{FOV}}}{H(m_s/2E_{\gamma} - 1)}\,,
\label{eq:diff_flux_EG}
\end{equation}
where $\rho_c$ is the present critical density, $\Omega_{\mathrm{dm}}\simeq 0.21$ is the dark
matter density parameter, and $H(z)$ is the Hubble function, where in this 
study we adopt a flat $\Lambda$-matter dominated universe with $H(0)=73~\mathrm{km}~\mathrm{s}^{-1}~\mathrm{Mpc}^{-1}$, as well as matter and dark energy density parameters equal to $\Omega_{\mathrm {m}} \simeq 0.24$ and $\Omega_\Lambda\simeq~0.76$ respectively~\cite{Spergel:2006hy}.

The sterile neutrinos in the Galactic halo give rise to the differential energy flux
\begin{equation}
\varphi_E^{\mathrm {MW}}= \frac{\Gamma_\gamma}{4\pi m_s}
\int_{\mathrm{FOV}}\int_{\mathrm{l.o.s.}}\rho_{\mathrm {dm}}(x, \Omega)~\mathrm{d}x~\mathrm{d}\Omega
\equiv\frac{\Gamma_\gamma}{4\pi m_s}\int_{\mathrm{FOV}}S_{\mathrm{dm}}(\Omega)~\mathrm{d}\Omega
\equiv\frac{\Gamma_\gamma}{4\pi m_s}\Omega_{\mathrm{FOV}}{\bar S_{\mathrm{dm}}},\;\label{eq:diff_flux_MW}
\end{equation}
which depends on the integral over the dark matter density $\rho_{\mathrm{dm}}$ along the line of sight (l.o.s.), $S_{\mathrm{dm.}}$.
\noindent Owing to the small variation of $S_{\mathrm{dm}}$ over the small FOV of \emph{Chandra} ($\sim$~5' circular area) in the directions of interest, we can approximately replace ${\bar S_{\mathrm{dm}}}$ (i.e. the average value of $S_{\mathrm{dm}}$ over the relevant FOV), with the value of $S_{\mathrm{dm}}$ at the centre of each FOV.

We calculate the flux contributions from decays within the Galactic halo using the Navarro-Frenk-White (NFW) density profile $\rho_{\mathrm{dm}}(r)=\rho_s~\left(r/r_s\right)^{-1}\left(1+\left(r/r_s\right)\right)^{-2}$ \cite{NFW},
where $\rho_s$ and $r_s$ are the scale density and radius respectively. We
adopt the recent evaluation of the virial mass of the MW halo, $0.6\times 10^{12}~M_{\odot}<M_{\mathrm{vir.}}<2.0\times 10^{12}~M_{\odot}$, from~\cite{Battaglia:2005rj}, which is also consistent with $R_{\mathrm{vir.}}=255$~kpc, $R_{\mathrm{vir.}}/r_s=18$ and $R_{\odot}=8$~kpc.

Adopting the above, we obtained values for the dark matter surface density of
$(S_{\mathrm{dm}})_{\mathrm{CDF-N}}=\left[0.0115, 0.0384\right]~\mathrm{g\,cm}^{-2}$ in the direction of CDF-N (with Galactic coordinates $(l, b)_{\mathrm{CDF-N}}=(125.89^{\circ}, 54.83^{\circ})$), and $(S_{\mathrm{dm}})_{\mathrm{CDF-S}}=\left[0.0111, 0.0369\right]~\mathrm{g\,cm}^{-2}$ in the direction of CDF-S (with Galactic coordinates $(l, b)_{\mathrm{CDF-S}}=(223.57^{\circ}, -54.44^{\circ})$), where the lower (higher) extremes of each the ranges stated correspond to the lowest (highest) values of the MW virial mass mentioned above.

\section{Results}
\begin{figure}[h]
   \hfill
   \begin{minipage}[t]{.50\textwidth}
     \begin{center}
       \includegraphics[width=70mm, keepaspectratio, clip]{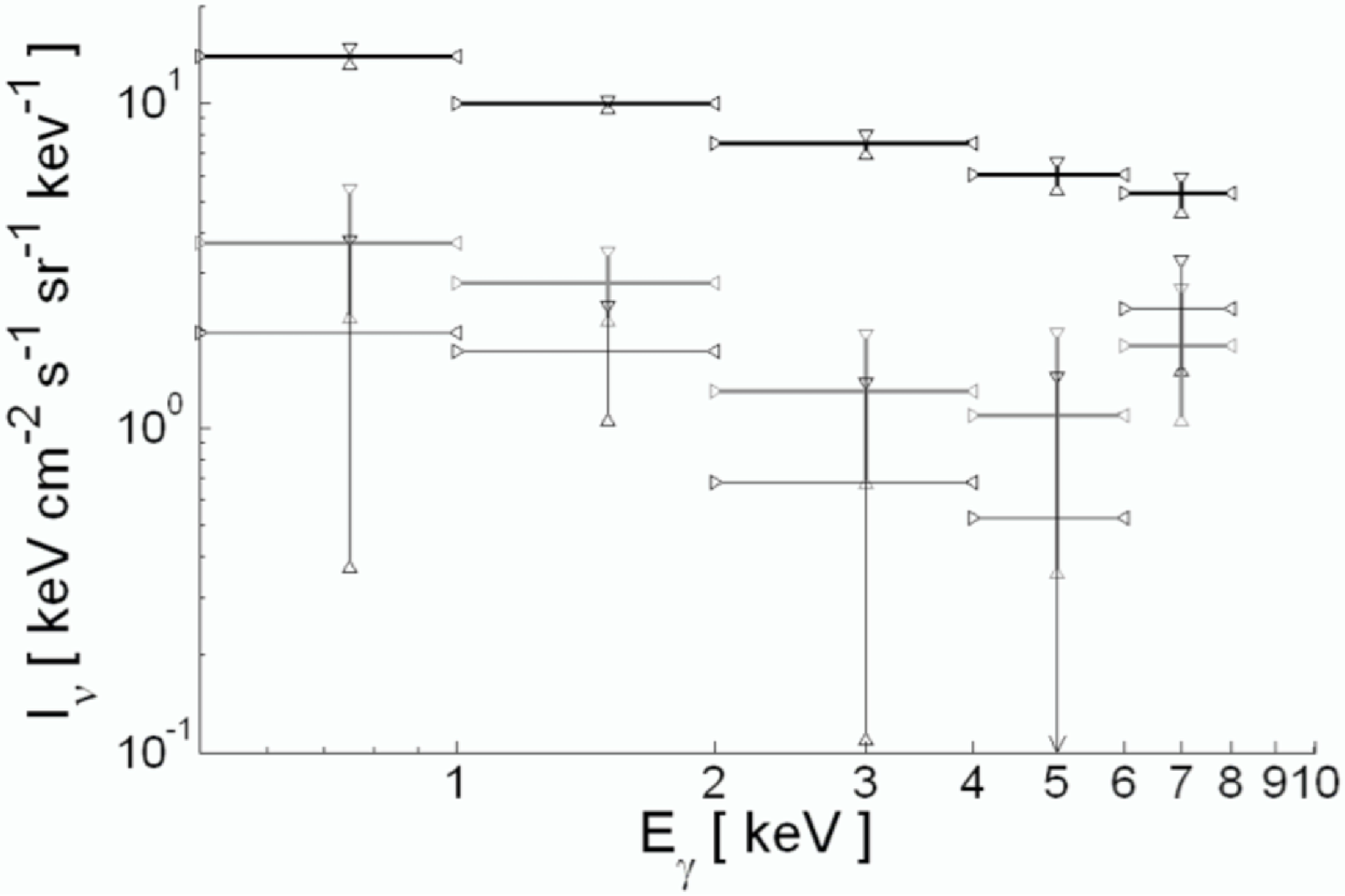}
     \end{center}
   \end{minipage}
   \hfill
   \begin{minipage}[t]{.50\textwidth}
     \begin{center}
        \includegraphics[width=70mm, keepaspectratio, clip]{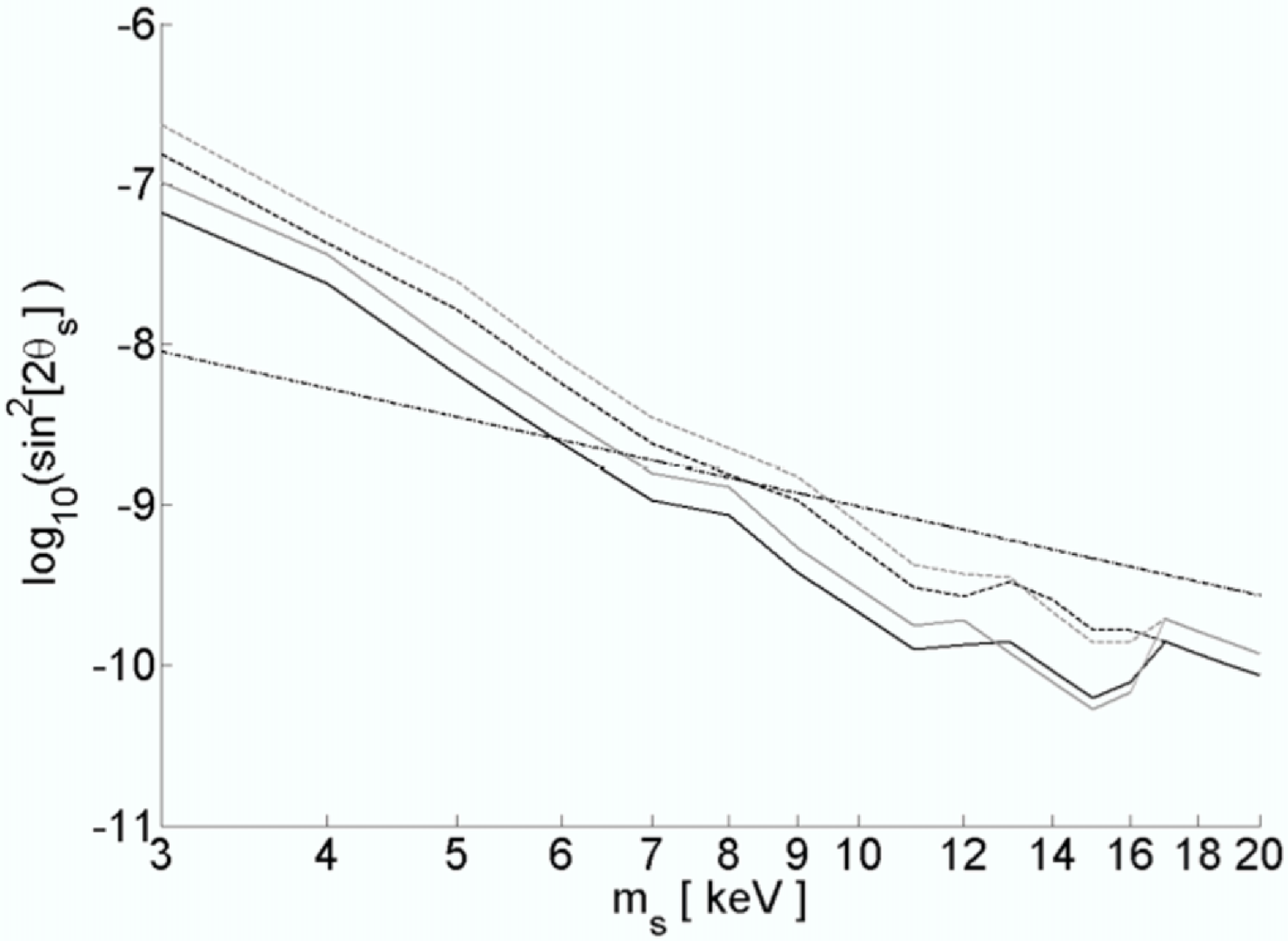}
       \caption{\emph{Left}\,-\,XRB spectra in terms of the total specific intensity $I_{\nu}=\mathrm{d}		    \varphi_E/\mathrm{d}\Omega$ ({\emph{thick-black crosses}}), and contributions from 		    unresolved sources ({\emph{thin-black crosses (CDF-N), thick-grey crosses (CDF-S)}} 		    ). The errors displayed are taken to be at the 1$\sigma$ level. \emph{Right}\,-\,1	             		   $\sigma$ upper limits on sin$^2(2\theta_s)$ using values of the Galactic halo virial 		   mass of $M_{\mathrm{vir.}}=0.6\times10^ {12}M_{\odot}$ (\emph{dashed lines}) and 	            $M_{\mathrm{vir.}}=0.6\times10^{12}M_{\odot}$ (\emph{solid lines}), for CDF-N (\emph		 {black lines}) and CDF-S (\emph{grey lines}). The \emph{dot-dashed line} indicates the 		 parameter space permitted for DW sterile neutrinos constituting all of the currently 			 inferred dark matter relic abundance.}
       \label{figure}
     \end{center}
   \end{minipage}
   
       \hfill
\end{figure}
Here, we utilise results from the analysis of the CDF observations by Worsley~\emph{et al.}~\cite{Worsley}. Fig.~\ref{figure} (\emph{left}) displays the \emph{total} XRB spectra aswell as the contributions attributed to unresolved sources. The errors displayed are taken to be at the 1$\sigma$ level. We then derive upper limits on the radiative decay rate of sterile neutrinos by applying the following spectral analysis.

For each selected value of $m_s$, we add the corresponding EG and MW contributions (after accounting for signal broadening, owing to \emph{Chandra's} finite energy resolution, (which we conservatively estimate as $\Delta E/E\ge0.03$~\cite{NASA}), and limit the decay rate \eqref{eq:Gamma} by invoking the following criterion

\begin{equation}
\int_{\Delta E_{\mathrm{dp}}} I_{\nu}^{CDF-N/S}~dE\ge\int_{\Delta E_{\mathrm{dp}}} 
\left(I_{\nu}^{\mathrm {EG}} +I_{\nu}^{\mathrm {MW}}\right)~dE 
\label{flux_criterion}
\end{equation}

\noindent where the left-hand integral is evaluated over the \emph{upper limit} of each of the data points displayed, each with energy range $\Delta E_{\mathrm{dp}}$. In fig.~\ref{figure} (\emph{right}), we display the resulting 1$\sigma$ upper limits on sin$^2(2\theta_s)$ as a function of $m_s$, obtained using values of the Galactic halo virial mass $M_{\mathrm{vir.}}=0.6\times10^{12}M_{\odot}$ (\emph{dashed lines}) and $M_{\mathrm{vir.}}=0.6\times10^{12}M_{\odot}$ (\emph{solid lines}). In addition, we plot a contour indicating the parameter space permitted for sterile neutrinos constituting all of the currently inferred dark matter relic abundance when generated solely via the DW mechanism (\emph{dot-dashed line}).

Hence, using fig.~\ref{figure} (\emph{right}), the 1$\sigma$ upper mass limits for DW sterile neutrino dark matter are
\[ m_s<\left\{ \begin{array}
	{r@{\quad\mathrm{keV\,\,\,for}\,\,\,}l}
   \left[5.95, 8.30\right] &  \mathrm{CDF-N\,\,\,\,observations}\\ 
   \left[6.65, 9.49\right] & \mathrm{CDF-S\,\,\,\,observations}
        \end{array} \right.	\]          
\noindent where once again, the lowest (highest) extremes of each mass range correspond to the lowest (highest) values mentioned above for the virial mass of the Galactic halo.

In conclusion, the above constraints are slightly less stringent than those obtained from other related works which utilise a line signal non-detection spectral analysis of the XRB data (see e.g. \cite{Abazajian:2006jc}). However, we consider that such techniques yield overly stringent results when the energy resolution of the data  is not sufficient enough to resolve line signals from local contributions. If so, then such results should be viewed as absolute lower/upper bounds on the sterile neutrino mass/mixing-angle, whereas the results obtained using the above \emph{integrated flux} method should be considered to be the converse bounds on these parameters.

\end{document}